\documentclass[twocolumn,prd,amsmath,superscriptaddress]{revtex4}
\usepackage{graphicx}
\usepackage{latexsym}
\usepackage{amsmath}
\usepackage{amssymb}
\usepackage{graphics}
\usepackage{color}
\usepackage{hyperref}
\usepackage{ifthen}
\usepackage{bm}
\usepackage{color}

\definecolor{darkgreen}{RGB}{0,180,0}

\newcommand\ee{\end{equation}}
\newcommand\be{\begin{equation}}
\newcommand\eea{\end{eqnarray}}
\newcommand\bea{\begin{eqnarray}}

\def\gsim{ \lower .75ex \hbox{$\sim$} \llap{\raise .27ex \hbox{$>$}} }
\def\lsim{ \lower .75ex \hbox{$\sim$} \llap{\raise .27ex \hbox{$<$}} }

\begin{document}

\title{Dark Matter Superfluidity and Galactic Dynamics}
\author{Lasha Berezhiani and Justin Khoury}
\affiliation{Center for Particle Cosmology, Department of Physics and Astronomy, University of Pennsylvania, Philadelphia, PA 19104, USA 
}


\begin{abstract}
We propose a unified framework that reconciles the stunning success of MOND on galactic scales with the triumph of the $\Lambda$CDM model
on cosmological scales. This is achieved through the physics of superfluidity. Dark matter consists of self-interacting axion-like particles that thermalize
and condense to form a superfluid in galaxies, with $\sim$mK critical temperature. The superfluid phonons mediate a MOND acceleration on baryonic matter. Our framework naturally distinguishes between galaxies (where MOND is successful) and galaxy clusters (where MOND is not): dark matter has a higher temperature in clusters, and hence is in a mixture of superfluid and normal phase. The rich and well-studied physics of superfluidity leads to a number of striking observational signatures.
\end{abstract}

\maketitle

The standard $\Lambda$ Cold Dark Matter ($\Lambda$CDM) model does very well at fitting large scale
observables. On galactic scales, however, a number of challenges have emerged. Disc galaxies display a tight
correlation between total baryonic mass and asymptotic velocity, $M_{\rm b} \sim v_{\rm c}^4$, known as the Baryonic Tully-Fisher
Relation (BTFR)~\cite{Freeman1999,McGaugh:2000sr}. Hydrodynamical simulations can reproduce the BTFR by tuning baryonic feedback processes,
but their stochastic nature naturally results in a much larger scatter~\cite{Vogelsberger:2014dza}. Furthermore, the mass~\cite{BoylanKolchin:2011de,toobigtofail2} and
phase-space~\cite{Pawlowski:2012vz,Pawlowski:2013cae,Ibata:2013rh,Ibata:2014pja} distributions of dwarf satellites in the Local Group are puzzling.

A radical alternative is MOdified Newtonian Dynamics (MOND)~\cite{Milgrom:1983ca}, which replaces dark matter (DM) with a modification of gravity at low acceleration: $a\simeq a_{\rm N}$ ($a_{\rm N} \gg a_0$); $a\simeq \sqrt{a_{\rm N}a_0}$ ($a_{\rm N} \ll a_0$), with best-fit value $a_0 \simeq 1.2\times 10^{-8}~{\rm cm}/{\rm s}^2$. This empirical force law has been remarkably successful at explaining a wide range of galactic phenomena~\cite{Famaey:2011kh}. In the MOND regime, a test particle orbits an isolated source according to $v^2/r = \sqrt{G_{\rm N} M_{\rm b}a_0/r^2}$. This gives a constant asymptotic velocity, $v_{\rm c}^2 = \sqrt{G_{\rm N} M_{\rm b} a_0}$, which in turn implies the BTFR. 

The empirical success of MOND, however, is limited to galaxies. The predicted temperature profile in galaxy clusters conflicts with observations~\cite{Aguirre:2001fj}. 
The Tensor-Vector-Scalar (TeVeS) relativistic extension~\cite{Bekenstein:2004ne} fails to reproduce the CMB and matter spectra~\cite{Skordis:2005xk,Zuntz:2010jp}. The lensing features of merging clusters~\cite{Clowe:2006eq,Harvey:2015hha} are problematic~\cite{Angus:2006qy}. This has motivated various hybrid proposals that include both DM and MOND, {\it e.g.},~\cite{Blanchet:2006yt,Bruneton:2008fk,Li:2009zzh,Ho:2010ca,Khoury:2014tka}. 

In this Letter, together with a longer companion paper~\cite{future}, we propose a novel framework that unifies the DM and MOND phenomena through the physics of superfluidity. Our central idea is that DM forms a superfluid inside galaxies, with a coherence length of order the size of galaxies. The critical temperature is $\sim {\rm mK}$, which intriguingly is comparable to Bose-Einstein condensation (BEC) critical temperatures for cold atom gases. Indeed, in many ways our DM behaves like cold dark atoms.

The superfluid nature of DM dramatically changes its macroscopic behavior in galaxies. Instead of evolving as independent particles, 
DM is more aptly described as collective excitations. Superfluid phonons, in particular, mediate a MOND-like force between baryons. Since superfluidity only occurs at low enough temperature,
our framework naturally distinguishes between galaxies (where MOND is successful) and galaxy clusters (where MOND is not). Due to the larger velocity dispersion in clusters,
DM has a higher temperature and hence is in a mixture of superfluid and normal phases~\cite{tisza,london,landau}.

The superfluid interpretation makes the non-analytic nature of the MOND scalar action more palatable. The Unitary Fermi Gas,
which has attracted much excitement in cold atom physics~\cite{UFGreview}, is also governed by a non-analytic kinetic term~\cite{Son:2005rv}.
Our equation of state $P\sim \rho^3$ suggests that the DM superfluid arises through three-body interactions. 
It would be fascinating to find precise cold atom systems with the same equation of state, as this would give important insights on the microphysical interactions underlying our superfluid.
Tantalizingly, this might allow laboratory simulations of galactic dynamics. 

The idea of DM BEC has been studied before~\cite{Sin:1992bg,Hu:2000ke,Goodman:2000tg,Peebles:2000yy,Boehmer:2007um}, with important differences from our work.
In BEC DM galactic dynamics are caused by the condensate density profile; in our case phonons play a key role in explaining the BTFR. 
Moreover, BEC DM has $P \sim \rho^2$ instead of~$\sim \rho^3$. This implies a much lower sound speed, which puts BEC DM in tension with observations~\cite{slepiangoodman}.

\vspace{0.2cm}
\noindent {\bf DM Condensation:} In order for DM particles to condense in galaxies, their de Broglie wavelength $\lambda_{\rm dB}\sim (mv)^{-1}$ must be larger than the interparticle separation $\ell\sim \left(m/\rho_{\rm vir}\right)^{1/3}$. From standard collapse theory,  the density at virialization is $\rho_{\rm vir} \simeq (1+z_{\rm vir})^3 \; 5.4\times 10^{-28}\; {\rm g}/{\rm cm}^3$, while the virial velocity is $v = 113\, M_{12}^{1/3}\sqrt{1 + z_{\rm vir}}~{\rm km}/{\rm s}$, where $M_{12}\equiv M/10^{12}M_\odot$. Thus $\lambda_{\rm dB} \;\gsim\; \ell$ implies
\be
m \;\lsim\; 2.3 \left(1+z_{\rm vir}\right)^{3/8}\;  M_{12}^{-1/4}\; {\rm eV}\,.
\ee

The second condition is that DM thermalizes, with temperature set by the virial velocity $v$. The interaction rate is
$\Gamma\sim {\cal N} v \rho_{\rm vir} \frac{\sigma}{m}$, where ${\cal N}  \sim \frac{\rho_{\rm vir}}{m} \frac{(2\pi)^3}{\frac{4\pi}{3}(mv)^3}$
is the Bose enhancement factor. The rate should be larger than the inverse dynamical time $t_{\rm dyn} \sim \frac{1}{\sqrt{G_{\rm N}\rho_{\rm vir}}}$, such that the coherence length will span the halo. This translates into a bound on the cross section (with $m_{\rm eV} \equiv m/{\rm eV}$):
\be
\sigma/m \;\gsim\;  \left(1+z_{\rm vir}\right)^{-7/2}m_{\rm eV}^{4}M_{12}^{2/3}\; 52~{\rm cm}^2/{\rm g}\,.
\label{siglow}
\ee
Later on, we will adopt $m = 0.6$~eV as a fiducial value. For $M_{12} = 1$ and $z_{\rm vir} = 2$, the inequality becomes
$\sigma/m \;\gsim\; 0.1\;{\rm cm}^2/{\rm g}$. The lower end is consistent with current constraints~\cite{MiraldaEscude:2000qt,Gnedin:2000ea,Randall:2007ph} on $\sigma/m$ for self-interacting dark matter (SIDM)~\cite{Spergel:1999mh}, though these constraints must be carefully revisited in the superfluid context. 

\begin{figure}[t]
\centering
\includegraphics[width=2.5in]{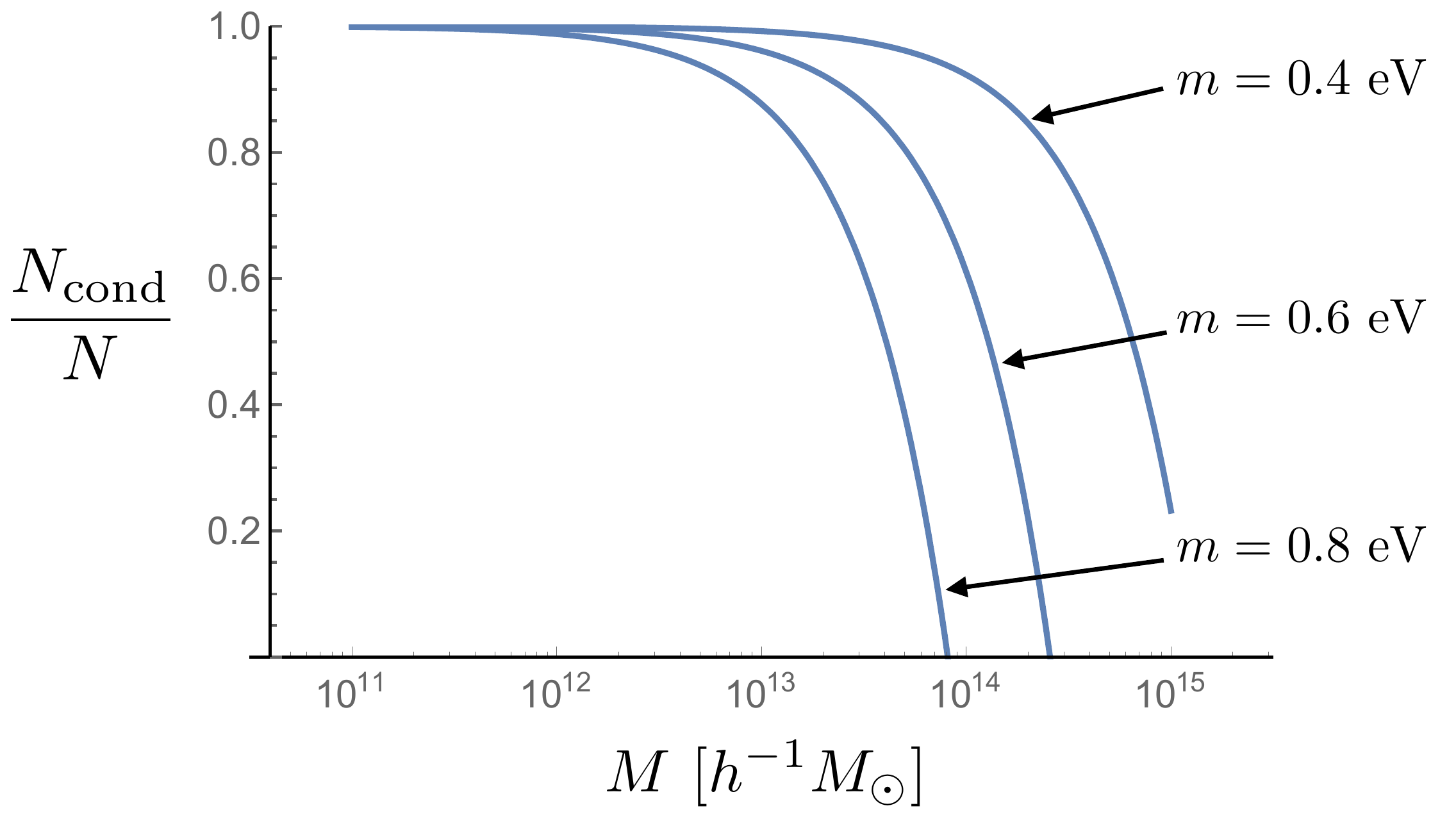}
\caption{\label{Ncond} \small Fraction of DM particles in the condensate.}
\end{figure}

The critical temperature, obtained by equipartition $k_{\rm B}T_{\rm c} = \frac{1}{3} mv^2_{\rm c}$,
is in the mK range:
\be
T_{\rm c} = 6.5 \; m_{\rm eV}^{-5/3} (1+z_{\rm vir})^{2}~{\rm mK}\,.
\ee
For $0 < T < T_{\rm c}$, the system is a mixture of condensate and normal components. The fraction of condensed particles, $1 - \left(T/T_{\rm c}\right)^{3/2}$~\cite{landaubook}, is shown in Fig.~\ref{Ncond} as a function of halo mass assuming $z_{\rm vir} = 0$. For $m\;\lsim\; {\rm eV}$, galaxies are almost completely condensed while massive clusters have a significant normal component. 

\vspace{0.2cm}
\noindent {\bf Superfluid Phase:} The relevant low-energy degrees of freedom of a superfluid are phonons, described by a scalar field $\theta$. 
In the presence of a gravitational potential $\Phi$, the non-relativistic effective action is ${\cal L} = P(X)$, where $X =  \dot{\theta} - m\Phi - (\vec{\nabla}\theta)^2/2m$~\cite{Son:2005rv}. We conjecture that DM superfluid phonons are governed by the MOND action~\cite{Bruneton:2007si}
\be
{\cal L} = \tfrac{2}{3} \Lambda(2m)^{3/2}X\sqrt{|X|} - \alpha\tfrac{\Lambda}{M_{\rm Pl}} \theta \rho_{\rm b}\,,
\label{PMOND}
\ee
where $\Lambda$ is a mass scale, $\rho_{\rm b}$ is the baryonic matter density, and $\alpha$ is a dimensionless constant.
This action should only be trusted {\it away from} $X = 0$, as we will see later. The matter coupling breaks
the shift symmetry at the $1/M_{\rm Pl}$ level and is thus technically natural. Remarkably~\eqref{PMOND}
is strikingly reminiscent of the Unitary Fermi Gas, ${\cal L}_{\rm UFG}(X) \sim X^{5/2}$, which is also non-analytic~\cite{Son:2005rv}.

The phonon action~\eqref{PMOND} uniquely fixes the properties of the condensate through standard thermodynamics.
At finite chemical potential, $\theta = \mu t$, and ignoring $\Phi$, the pressure is given by the Lagrangian density:
\be
P(\mu) = \tfrac{2}{3}\Lambda (2m\mu)^{3/2}\,.
\label{Pmu}
\ee
This is the grand canonical equation of state $P = P(\mu)$ for the condensate. The number density, $n =  \partial P/\partial\mu$,
is
\be
n = \Lambda (2m)^{3/2} \mu^{1/2}\,.
\label{nmu}
\ee
Combining these expressions with $\rho = m n$, we obtain
\be
P = \tfrac{\rho^3}{12\Lambda^2m^6}\,.
\label{eos}
\ee
This is a polytropic equation of state $P \sim \rho^{1 + 1/n}$ with index $n = 1/2$. In comparison, BEC DM has $P\sim \rho^2$~\cite{Goodman:2000tg}. 

Including phonons excitations $\theta = \mu t + \phi$, the quadratic action for $\phi$ is  ${\cal L}_{\rm quad} =\frac{\Lambda (2m)^{3/2}}{4\mu^{1/2}} \left( \dot{\phi}^2 - \frac{2\mu}{m} (\vec{\nabla}\phi)^2\right)$. The sound speed can be immediately read off:
\be
c_s = \sqrt{2\mu/m}\,.
\label{cs}
\ee

\begin{figure}[t]
\centering
\includegraphics[width=2.in]{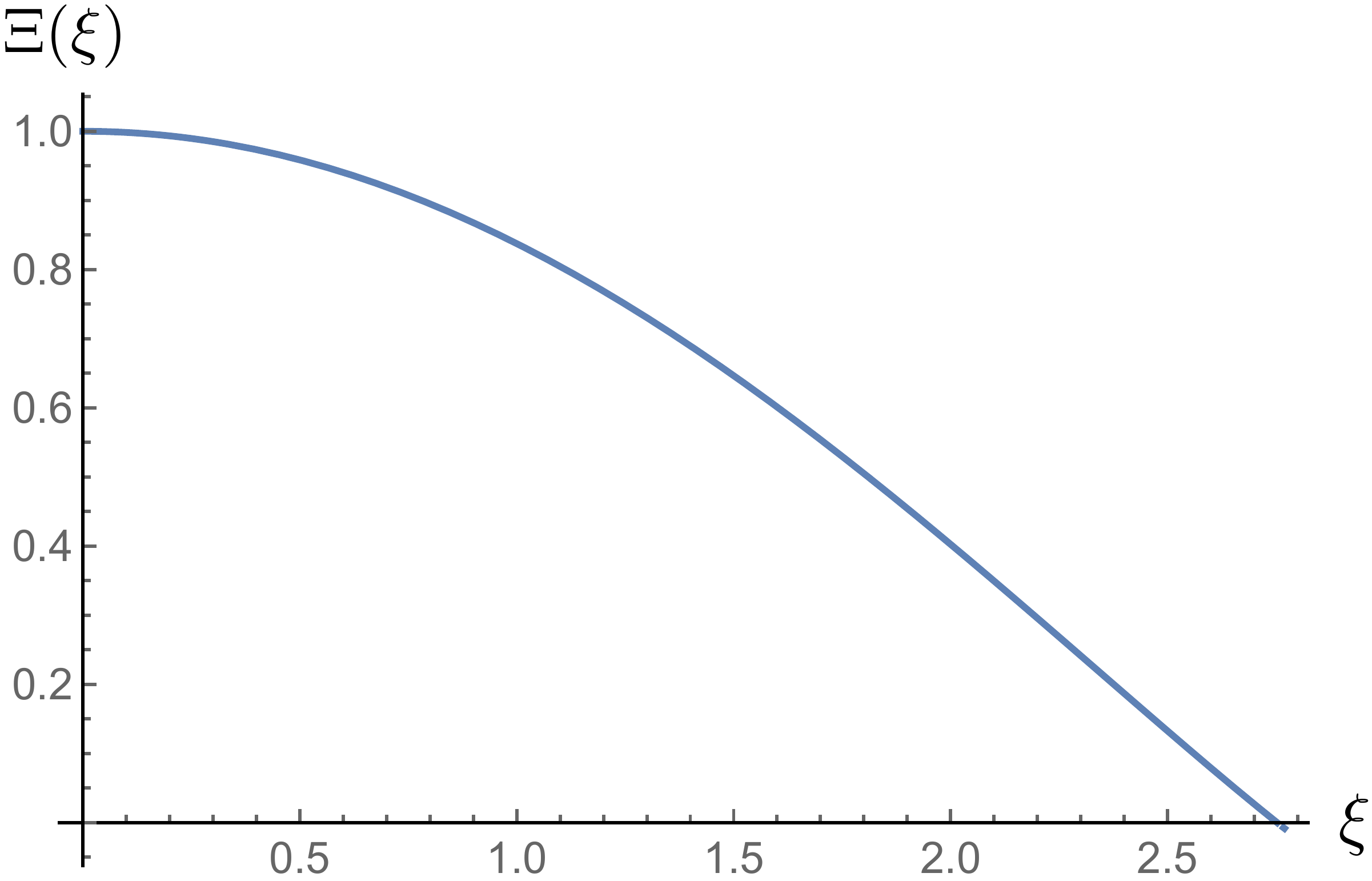}
\caption{\label{laneemden} \small Numerical solution of Lane-Emden equation~\eqref{LaneEmdenEqn}.}
\end{figure}

Using~\eqref{eos}, we compute the static, spherically-symmetric density profile of the DM condensate halo. Introducing dimensionless variables $\rho = \rho_0\Xi$ and $r = \sqrt{\frac{\rho_0}{32\pi G_{\rm N} \Lambda^2 m^6}}\,\xi$, with $\rho_0 $ denoting the central density, hydrostatic equilibrium implies the Lane-Emden equation,
\be
\left(\xi^2 \Xi'\right)' = - \xi^2 \Xi^{1/2}\,,
\label{LaneEmdenEqn}
\ee
with boundary conditions $\Xi(0) = 1$ and $\Xi'(0) = 0$. The numerical solution is shown in Fig.~\ref{laneemden}.
It vanishes at $\xi_1 \simeq 2.75$, which defines the halo size: $R = \sqrt{\frac{\rho_0}{32\pi G_{\rm N} \Lambda^2 m^6}}\; \xi_1$.
Meanwhile the central density is related to the halo mass as~\cite{chandrabook} $\rho_0 =  \frac{3M}{4\pi R^3} \frac{\xi_1}{|\Xi'(\xi_1)|}$,
with $\Xi'(\xi_1)\simeq -0.5$. Combining these results, we can solve for $\rho_0$ and $R$:
\bea
\nonumber
\rho_0 &\simeq & M_{12}^{2/5} m_{\rm eV}^{18/5}\Lambda_{\rm meV}^{6/5} \; 7 \times 10^{-25}~{\rm g}/{\rm cm}^3  \,; \\
R &\simeq & M_{12}^{1/5} m_{\rm eV}^{-6/5}\Lambda_{\rm meV}^{-2/5} \; 36~{\rm kpc}\,,
\label{halorad}
\eea
where $\Lambda_{\rm meV}\equiv \Lambda/{\rm meV}$. Remarkably, with $m\sim {\rm eV}$ and $\Lambda\sim {\rm meV}$
we obtain DM halos of realistic size. Concretely, as fiducial values we will fix
\be
m = 0.6~{\rm eV}\,;\qquad \Lambda = 0.2~{\rm meV}\,.
\label{fidparam}
\ee
This implies $R \sim 125$~kpc for $M_{\rm DM} = 10^{12}\,M_\odot$. In general, we expect this superfluid core to be surrounded by a cloud of DM particles in the normal phase, likely described by a Navarro-Frenk-White profile~\cite{Navarro:1996gj}.

\vspace{0.2cm}
\noindent {\bf Including Baryons:} We now derive the phonon profile $\theta = \mu t + \phi(r)$ in galaxies, modeling baryons as a static, spherically-symmetric source. The equation of motion, $\vec{\nabla} \cdot \left( \sqrt{2m|X|}~\vec{ \nabla}\phi\right) = \alpha\rho_{\rm b}(r)/2M_{\rm Pl}$, 
where $X = \mu -m\Phi(r) - \phi'^2/2m$, can be readily integrated:
\be
\sqrt{2m|X|}~\phi' = \frac{\alpha M_{\rm b}(r)}{8\pi M_{\rm Pl} r^2}\equiv \kappa(r)\,.
\label{kappadef}
\ee
There are two branches of solutions, depending on the sign of $X$. We focus on the MOND branch (with $X < 0$):
\be
\phi' (r) = \sqrt{m}\left( \hat{\mu}  + \sqrt{\hat{\mu}^2 + \kappa^2/m^2}\right)^{1/2} \,,
\label{phigensoln}
\ee
where $\hat{\mu} \equiv \mu - m\Phi$. Indeed, for $\kappa/m \gg \hat{\mu}$ we have
\be
\phi'(r) \simeq  \sqrt{\kappa(r)} \,.
\label{phiMOND}
\ee
In this limit the scalar acceleration on baryonic particle is $a_\phi(r) = \alpha \Lambda \phi'/M_{\rm Pl}$. Matching to the MOND result $a_{\rm MOND}  = \sqrt{a_0a_{\rm N}}$ fixes $\alpha$ in terms of $\Lambda$ and $a_0$:
\be
\alpha^{3/2} \Lambda = \sqrt{a_0 M_{\rm Pl}}\simeq 0.8~{\rm meV}\,.
\label{alphasoln}
\ee
For the fiducial value~\eqref{fidparam} $\Lambda = 0.2$~meV, we get $\alpha\simeq 2.5$.

As it stands, the $X < 0$ solution is unstable. It leads to unphysical halos, with growing DM density profiles.
The instability can be seen by expanding~\eqref{PMOND} in $\varphi = \phi - \bar{\phi}(r)$ to obtain ${\cal L} ={\rm sign}(\bar{X}) \frac{\Lambda(2m)^{3/2}}{4\sqrt{|\bar{X}|}} \dot{\varphi}^2 + \ldots$. Clearly the kinetic term is ghostly for $\bar{X} < 0$. (The $X > 0$ branch is stable but does not admit a MOND regime~\cite{future}.) 

Since DM in actual galactic halos has $T\neq 0$, however, we expect~\eqref{PMOND} to
receive finite-temperature corrections in galaxies. At finite sub-critical temperature, the Lagrangian depends on $X$ and
two additional scalars~\cite{Nicolis:2011cs}
\be
B \equiv {\rm det}^{1/2}\partial_\mu\psi^I\partial^\mu\psi^J \,;~~  Y \equiv  \hat{\mu} +  \dot{\phi} + \vec{v}\cdot \vec{\nabla}\phi\,,
\label{BY}
\ee
where $\psi^I(\vec{x},t)$, $I = 1,2,3$, and $\vec{v}$ are respectively the Lagrangian coordinates and velocity of the normal fluid.
For instance, consider the 2-derivative operator
\be
\Delta {\cal L} = M^2(T) Y^2 \simeq M^2(\hat{\mu} +  \dot{\phi})^2\,, 
\ee
where we have specialized to the rest frame of the normal fluid, $\vec{v} = 0$.
This leaves the static profile~\eqref{phigensoln} unchanged, but modifies the quadratic Lagrangian by $M^2 \dot{\varphi}^2$,
restoring stability for sufficiently large $M$. Specifically this is the case for $M\;\gsim\; {\rm eV}$~\cite{future}, remarkably of the same order as $m$! 
By the same token it corrects the condensate pressure by an amount $\Delta P = M^2 \mu^2$, which obliterates the unwanted
growth in the DM density profile, resulting in localized, finite-mass halos.

Another possibility is the finite-T Lagrangian
\be
P(X,T) =   \tfrac{2}{3}\Lambda(2m)^{3/2}X\sqrt{\left\vert X- \beta(T) Y \right\vert}\,.
\label{LphonT}
\ee
where $\beta > \frac{3}{2}$, as required for stability~\cite{future}. The DM condensate pressure is $P(\mu,T) = \frac{2}{3}\sqrt{\beta -1}\Lambda (2m\mu)^{3/2}$, which is identical to~\eqref{Pmu} modulo the redefinition $\Lambda\rightarrow \sqrt{\beta-1} \Lambda$. The halo density profile is therefore identical.

Including baryons, it is easy to show that~\eqref{kappadef} becomes
\be
\frac{\phi'^2 + 2m\left(\tfrac{2}{3}\beta -1\right)\hat{\mu}}{\sqrt{\phi'^2 + 2m(\beta -1)\hat{\mu}}}~\phi' = \kappa(r) \,.
\label{neweom}
\ee
At small distances ($\phi'^2 \gg m\hat{\mu}$), the solution approximates the MOND profile $\phi'\simeq \sqrt{\kappa}\sim 1/r$, and this gives the dominant force on a test baryonic particle~\cite{future}. At large distances ($\phi'^2 \ll m\hat{\mu}$), the solution tends to $\phi'\sim \kappa$, but this is subdominant to the gravitational acceleration due to the DM halo. The transition radius $r_\star$ delineating the MOND regime ($r\ll r_\star$) from the DM-halo regime ($r\gg r_\star$) occurs when $ \kappa \sim m\hat{\mu}$. Substituting~\eqref{kappadef} and approximating $\hat{\mu}$ by its central value $\rho^2_0/8\Lambda^2m^5$, we find
\be
r_{\rm t} \sim M_{11,\,{\rm b}}^{1/10}\left(\tfrac{M_{\rm b}}{M_{\rm DM}}\right)^{2/5} 
m_{\rm eV}^{-8/5}\Lambda_{\rm meV}^{-8/15}\, 28\;{\rm kpc}\,,
\ee
where $M_{11,\,{\rm b}}\equiv M_{\rm b}/10^{11}\,M_\odot$. For a galaxy with $M_{11,\,{\rm b}} = 3$ and cosmic DM-baryon ratio
$M_{\rm DM}/M_{\rm b} = \Omega_{\rm DM}/\Omega_{\rm b} \simeq 6$, the MOND regime extends to $r_\star \sim 70\;{\rm kpc}$.

\vspace{0.2cm}
\noindent {\bf Relativistic Completion:} It is well-known that a superfluid is described in the weak-coupling regime as a self-interacting complex scalar field with global $U(1)$ symmetry. Consider the relativistic field theory
\be
{\cal L} = -|\partial \Phi |^2 - m^2 |\Phi |^2 -  \tfrac{\Lambda^4 \left(|\partial \Phi |^2+ m^2 |\Phi |^2\right)^3}{3\left(\Lambda_{\rm c}^2 + |\Phi |^{2}\right)^6}\,,
\ee
where $\Lambda_{\rm c}$ makes $\Phi = 0$ well-defined. Substituting $\Phi= \rho  e^{i(\theta + mt)}$ and taking the non-relativistic limit, we obtain
\be
{\cal L} = - (\vec{\nabla}\rho)^2 + 2m\rho^2 X - \tfrac{\Lambda^4\left((\vec{\nabla}\rho)^2 - 2m\rho^2 X\right)^3}{6\left(\Lambda_{\rm c}^2 + \rho^2\right)^6} \,.
\label{Lwithrho}
\ee
To leading order in gradients we ignore $(\vec{\nabla}\rho)^2$ contributions and integrate out $\rho$. In the limit $\rho \gg \Lambda_{\rm c}$ this gives
\be
\rho^2 \simeq  \Lambda \sqrt{2m} \left(X^2\right)^{1/4} = \Lambda \sqrt{2m|X|} \,.
\label{rhosoln}
\ee
Substituting this back into~\eqref{Lwithrho}, we recover the MOND action (given by the kinetic part of~\eqref{PMOND}) for $\rho \gg \Lambda_{\rm c}$.

A finite $\Lambda_{\rm c}$ implies that MOND is restricted to $\rho \;\gsim\; \Lambda_{\rm c}$,
{\it i.e.}, $a_\phi \;\gsim\;  \frac{\Lambda_{\rm c}}{\alpha^2\Lambda} \; a_0$. Thus the MOND regime does not apply to arbitrarily small accelerations. By choosing $\Lambda_{\rm c}$ a factor of a few smaller than $\Lambda$, the predicted departure from MOND can occur around the acceleration scale of the Milky Way dwarf spheroidals, which are well-known to pose a challenge for MOND~\cite{Spergel,Milgrom:1995hz,Angus:2008vs,Hernandez:2009by,McGaugh:2010yr,Lughausen:2014hxa}.

\vspace{0.2cm}
\noindent {\bf Cosmology:} Given their mass, our DM particles are axion-like. The simplest genesis scenario is through
vacuum displacement, with DM being generated at a time when $H_i\sim m \sim {\rm eV}$. This corresponds to a
photon temperature of $\sim$TeV,  {\it i.e.} around the weak scale! The DM rapidly reaches thermal equilibrium with
itself and becomes superfluid, but is decoupled from ordinary matter.

Naturally DM is much colder cosmologically than in collapsed structures. The ratio $\frac{T}{T_{\rm c}} = \left(\frac{vm^{4/3}}{\rho^{1/3}}\right)^2$, which is constant cosmologically,
can be evaluated at matter-radiation equality using $\rho_{\rm eq} \simeq 0.4~{\rm eV}^4$ and $v_{\rm eq}  = v_i \frac{a_i}{a_{\rm eq}} \simeq \frac{{\rm eV}}{\sqrt{mM_{\rm Pl}}}$. The result is $\frac{T}{T_{\rm c}} \simeq 10^{-28}$, which is much colder than the typical range $10^{-6}-10^{-2}$ in galaxies. To obtain an acceptable cosmology, we need $\Lambda$ and $\alpha$ to assume different values cosmologically. We will denote their cosmological
values by $\Lambda_0$ and $\alpha_0$. 

On a cosmological background, the phonon equation of motion that derives from~\eqref{PMOND} is given by
\be
\tfrac{{\rm d}}{{\rm d}t} \left((2m)^{3/2}a^3\dot{\theta}^{1/2}\right) =  -\tfrac{\alpha_0}{M_{\rm Pl}}a^3\rho_{\rm b}\,.
\ee
Since $a^3\rho_{\rm b} = {\rm const.}$, this can be integrated straightforwardly.
The resulting (non-relativistic) energy density $\rho = m n  =  m\Lambda_0 (2m)^{3/2}\dot{\theta}^{1/2}$ is given by
\be
\rho =  -t\frac{\alpha_0\Lambda_0}{M_{\rm Pl}}mt \rho_{\rm b}  +  \rho_{\rm dust}\,,
\ee
where $\rho_{\rm dust}\sim a^{-3}$. In the matter-dominated era, the baryonic contribution $\sim \rho_{\rm b} t$ redshifts as $1/a^{3/2}$.
In order for the superfluid to behave as ordinary dust, the second term should dominate over the first all the way to the present time:
$\rho_{\rm dust}\;\gsim\; \frac{\alpha_0\Lambda_0}{M_{\rm Pl}}mt_0 \rho_{\rm b}$. Substituting the age of the universe
$t_0 = 13.9\times 10^9~{\rm yrs}$ and assuming a DM-to-baryon ratio of $\rho_{\rm dust}/\rho_{\rm b} = 6$, we obtain (with $m\sim {\rm eV}$)
\be
\alpha_0 \;\lsim\; 2.4 \times 10^{-5} \; \tfrac{{\rm eV}}{\Lambda_0}  \,.
\label{al0}
\ee

Since $\dot{\theta} \simeq \frac{\rho_{\rm dust}^2}{8m^5\Lambda_0} \sim \frac{1}{a^3}$, the phonon
velocity increases with redshift and inevitably results in a breakdown of the non-relativistic
approximation, $\dot{\theta}\ll m$. In other words, the superfluid becomes relativistic at sufficiently high density.
This is consistent with the equation of state~\eqref{eos}: $w = \frac{P}{\rho} = \frac{\rho^2}{12\Lambda^2_0m^6}$.
Demanding that it be non-relativistic by matter-radiation equality puts a lower bound on $\Lambda_0$ 
\be
\Lambda_0\;\gsim\; 0.1 \;{\rm eV}\,.
\label{L0}
\ee
This is roughly four orders of magnitude larger than the fiducial value $\Lambda = 0.2$~eV
assumed in galaxies. This can be achieved, for instance, if $\Lambda$ depends on temperature as
$\Lambda(T) =  \frac{\Lambda_0}{1 + \kappa_\Lambda (T/T_{\rm c})^{1/4}}$, with $\kappa_\Lambda \sim 10^4$.
Meanwhile,~\eqref{al0} implies $\alpha_0 \;\lsim\; 10^{-4}$, which is roughly four orders of magnitude smaller
than the ${\cal O}(1)$ value obtained in galaxies. This can be achieved, for instance, if 
$\alpha(T) =  \alpha_0\left(1 + \kappa_\alpha (T/T_{\rm c})^{1/4}\right)$, with $\kappa_\alpha \sim 10^4$.
Note that the scale $\Lambda' \sim \alpha \Lambda$ appearing in the phonon-baryon
coupling in~\eqref{PMOND} is nearly temperature-independent. 

\vspace{0.2cm}
\noindent {\bf Gravitational Lensing:} In TeVeS~\cite{Bekenstein:2004ne} the complete absence of DM requires introducing a time-like vector field $A_\mu$,
as well as a complicated coupling between $\phi$, $A_\mu$ and baryons in order to reproduce lensing observations. In our case, there is no need to introduce an extra vector,
as the normal fluid already provides us with a time-like vector $u^\mu$. Moreover, our DM contributes to lensing, so we are free to generalize the TeVeS coupling.

Suppose matter fields couple to the effective metric
\be
\tilde{g}_{\mu\nu} \simeq g_{\mu\nu} - 2\alpha\tfrac{\Lambda}{M_{\rm Pl}} \phi (\gamma g_{\mu\nu} +(1 + \gamma) u_\mu u_\nu)\,,
\label{ourg}
\ee
with $\gamma= 1$ corresponding to TeVeS. In the weak-field limit, this gives $h_{00} = -1 -  2(\Phi + \alpha\Lambda \phi/M_{\rm Pl})$, $h_{ij} = \delta_{ij} - 2(\Phi  + \gamma\alpha\Lambda \phi/M_{\rm Pl})\delta_{ij}$, where $\nabla^2\Phi = 4\pi G_{\rm N} \left(\rho_{\rm b} + \rho_{\rm DM}\right)$. Hence the lensing signal arises from a combination of the $u_\mu u_\nu$ disformal coupling and the DM condensate density profile. Determining the allowed range of $\gamma$ will require a detailed study, which is beyond the scope of this paper. What is clear is that there should be considerably more freedom than in TeVeS. It may even be the case
that $\gamma = -1$ is allowed, such that the coupling to matter would reduce to a simple conformal coupling.

\vspace{0.2cm}
\noindent {\bf Observational Implications:} We conclude with some astrophysical implications of our DM superfluid.

\noindent {\it Vortices}: When spun faster than a critical velocity, a superfluid develops vortices. The typical angular velocity of halos is well
above critical~\cite{future}, giving rise to an array of DM vortices permeating the disc~\cite{Silverman:2002qx}. It will be interesting to see whether these vortices
can be detected through substructure lensing, {\it e.g.}, with ALMA~\cite{Hezaveh:2012ai}.

\noindent {\it Galaxy mergers}: A key difference with $\Lambda$CDM is the merger rate of galaxies. Applying Landau's criterion, we find two possible outcomes.
If the infall velocity $v_{\rm inf}$ is less than the phonon sound speed $c_s$ (of order the viral velocity~\cite{future}),
then halos will pass through each other with negligible dissipation, resulting in multiple encounters and a longer merger time. If $v_{\rm inf} \;\gsim\; c_{\rm s}$, however, the
encounter will excite DM particles out of the condensate, resulting in dynamical friction and rapid merger.  

\noindent {\it Bullet Cluster}: For merging galaxy clusters, the outcome also depends on the relative fraction of superfluid vs normal components in the clusters.
For subsonic mergers, the superfluid cores should pass through each other with negligible friction (consistent with the Bullet Cluster),
while the normal components should be slowed down by self interactions. Remarkably this picture is consistent with the lensing map of the Abell 520 
``train wreck"~\cite{Mahdavi:2007yp,Jee:2012sr,Clowe:2012am,Jee:2014hja}, which show lensing peaks coincident with galaxies (superfluid components),
as well as peaks coincident with the X-ray luminosity peaks (normal components).

\noindent {\it Dark-bright solitons:} Galaxies in the process of merging should exhibit interference patterns (so-called dark-bright solitons) that have been observed in BECs counterflowing at super-critical velocities~\cite{exptpaper}. This can potentially offer an alternative mechanism to generate the spectacular shells seen around elliptical galaxies~\cite{shells}.

\noindent {\it Globular clusters:} Globular clusters are well-known to contain negligible amount of DM, and as such pose a problem for MOND~\cite{Ibata:2011ri}. In our case the presence of a significant DM component is necessary for MOND. If whatever mechanism responsible for DM removal in $\Lambda$CDM is also effective here, our model would predict DM-free (and hence MOND-free) globular clusters.

\vspace{0.1cm}
\noindent{\em Acknowledgments.} We thank A.~Arvanitaki, L.~Blanchet, A.~Erickcek, B.~Famaey, L.~Hui, B.~Jain, R.~Kamien, A.~Kosowsky, T.~Lubensky, S.~McGaugh, A.~Nicolis, M.~Pawlowski, J.~Peebles, R.~Sheth, D.~Spergel, P.~Steinhardt and M.~Zaldarriaga. J.K. is supported by NSF CAREER Award PHY-1145525 and NASA ATP grant NNX11AI95G. L.B. is supported by funds provided by the University of Pennsylvania.

\pagebreak


\end{document}